\documentclass[a4paper, amsfonts, amssymb, amsmath, reprint]{revtex4-2}

\usepackage{graphicx}
\usepackage{dcolumn}
\usepackage{bm}

\usepackage[utf8]{inputenc}
\usepackage[T1]{fontenc}
\usepackage{mathptmx}

\usepackage{amsmath}
\usepackage{amssymb}

\usepackage{xcolor}

\renewcommand{\rm}[1]{\mathrm{#1}}

\newcommand{\fref}{Fig.\,\ref}

\begin{document}

\title{Efficient timing jitter simulation for passively mode-locked semiconductor lasers} 

\author{Stefan Meinecke}
\email[]{meinecke@tu-berlin.de}

\author{Kathy Lüdge}
\affiliation{Institut für Theoretische Physik, Technische Universität Berlin, Hardenbergstr. 36, 10623 Berlin, Germany}


\begin{abstract}
Efficient simulation of the timing jitter in passively mode-locking lasers is key to their numerical investigation and optimization. We introduce a method based on the pulse-period fluctuation auto-correlation function and compare it against established methods with respect to their estimate error. Potential improvements of the computational cost by about two orders of magnitude are reported.
This advantage may facilitate larger parameter studies of passively mode-locked laser on small scale clusters or even desktop computers and thereby guide the target-oriented design of future lasers with ultra low timing jitter. 
\end{abstract}

\maketitle 


Passively mode-locked semiconductor lasers emit sequences of short equidistant optical pulses at high repetition rates without external driving \cite{HAU00,AVR00}. They find applications in science and technology ranging from optical communication \cite{RAF07, BIM06} to metrology \cite{UDE02, KEL03}, medical imaging \cite{LOE96} and optical clocking \cite{DEL91}.
An excellent temporal stability of the pulse positions, i.e. a small timing jitter, is crucial for many applications \cite{JIA02b,VAL07} and thus determines the usefulness of a given device.
The experimental characterization is typically performed by one of two spectral methods: The von der Linde method \cite{LIN86,KOL86} estimates the timing jitter by integrating the noise side-bands of the harmonics. It was originally developed for active mode-locking, but when carefully applied can also be used for passive mode-locking \cite{HAU93a, PAS04a, MAN14a, OTT14b}. The Kéfélian method \cite{KEF08} on the other hand was specifically developed for passive mode-locking and estimates the timing jitter from a Lorentzian shaped \cite{HAU93a,ELI97} repetition rate linewidth. As higher harmonics are not required, it is especially suited for the characterization of monolithically integrated laser with high repetition rates \cite{BRE10,LIN11f,DRZ13a,AUT19}.

In the process of understanding, optimizing and designing passively mode-locked lasers, models and simulations are vital tools \cite{JAV11, ROS11e, OTT14a, JAU16, BAR18, MEI19, HAU19}. However, the direct estimation of the timing jitter from either simulated spectra or sequences of pulse positions \cite{LEE02c,MUL06,OTT14a} often poses the problem of excessively large computational costs, as large numbers of simulated pulses are required. In the case of weak noise sources, analytical approaches have been developed for the Haus master equation framework \cite{HAU93a, JIA01} and semi-analytical approaches for delay-differential equation descriptions \cite{PIM14b,JAU15}.

In this manuscript, we study a three section tapered quantum dot laser \cite{MEI19}, for which the timing jitter analysis can neither be performed in the analytical Haus or in the semi-analytical delay-differential equation formalism, since the noise induced perturbations are too large to be treated linearly. Since the simulation of noisy time series dominates the overall computational cost, the goal is to obtain a good timing jitter estimate from a sample set as small as possible. We therefore present a brief overview of the established calculation methods and introduce a method, which is based on the auto-correlation function of the pulse-period fluctuations. We subsequently compare the standard errors of the various methods with respect to the sample set size. We find that the Kéfélian and our auto-correlation based method vastly outperform the other methods and henceforth discourage computing the timing jitter directly from pulse position.


The timing jitter analysis presented in this manuscript is performed on a theoretical description of a three section tapered quantum dot laser. The laser contains 10 layers of InAs quantum dots and the 3\,mm long cavity consists of a 0.7\,mm straight section, a 0.7\,mm absorber section and a 1.6\,mm tapered section with a full taper angle of 2°. The tapered output facet is anti-reflection coated while the opposing facet is high-reflection coated.
The model couples a traveling-wave equation for the electric field propagation to a semi-classical multi-population carrier model. 
The modelling hierarchy as well as the model parameters are chosen to best produce experimental results. The details can be found in \cite{MEI19}. 

\begin{figure}[htbp]
\centering
\includegraphics[width=\linewidth]{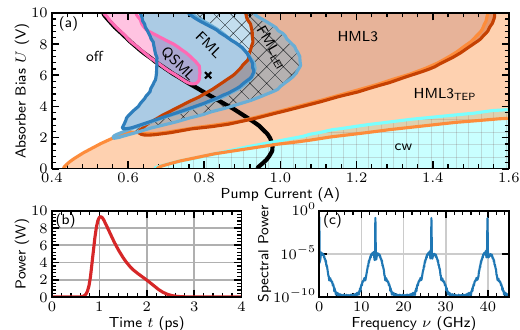}
\caption{Mode-locking characteristics: (a) Stable regions of the off, cw and the fundamental (FML), third order harmonic (HML3) and Q-switched mode-locking (QSML) states. LEI denotes a leading edge instability and TEP a strong trailing edge plateau. The solid black line represents the lasing threshold. The black cross indicates the parameters used for further investigations: (b) Pulse shape. (c) Normalized power spectral density.}
\label{fig:Laser}
\end{figure}

To characterize the emission dynamics, \fref{fig:Laser}\,(a) presents the stable regions of the off, cw state and the fundamental (FML), third order harmonic (HML3) and Q-switched mode-locking (QSML) states in the parameter space of the pump current $P$ and the absorber reverse bias $U$ as found by direct integration. The subscripts LEI and TEP denote either a leading edge instability or strong trailing edge plateau\cite{RAD11a}. Overlapping regions indicate a multistability between various states.
The subthreshold multistability observed for reverse biases below $U\approx 6.8$\,V can manifest as a hysteresis around the lasing threshold\cite{ERN10b, AUT19a}.

Stable fundamental mode-locking (blue), on which the further jitter analysis focuses, is only observed in region for reverse biases above $U \approx 2.6$\,V and for pump currents between $P \approx 0.55$\,A and $ P\approx 0.95$\,A. At lower reverse biases the FML state can be bistable with either the off, QSML or HML3 state, while at larger reverse biases multistabilities are first reduced and then disappear.

We chose $P=0.81$\,A and $U=6$\,V, indicated by a black cross in \fref{fig:Laser}\,(a) as the operating point for our jitter analysis. The respective dynamics are illustrated by a time series slice in \fref{fig:Laser}\,(b), portraying the pulse shape with $\approx 10$\,W peak power and $\approx 550$\,fs pulse width and the power spectral density in \fref{fig:Laser}\,(c). The first peak in the spectrum represents the fundamental repetition rate of $\approx13.24$\,GHz, which corresponds to a pulse round-trip time of $\approx75.16$\,ps.


On sufficiently large time scales, the pulse positions of passively mode-locked lasers perform a random walk with respect to an ideal jitter free pulse train \cite{HAU93a,ELI97}. Therefore, the respective diffusion coefficient $D_{\Delta t}$ represents an unambiguous measure to characterize the long term timing stability. This manuscript in particular uses the long term timing jitter $\sigma_{\mathrm{lt}}$, which relates to the diffusion coefficient via $\sigma_{\mathrm{lt}} = \sqrt{2 D_{\Delta t}}$. The following paragraphs present the time domain definition of $\sigma_{\mathrm{lt}}$ and introduce a method, which is based on the auto-correlation function of the pulse period fluctuations. Details on the spectral von der Linde\cite{LIN86} and Kéfélian\cite{KEF08} methods can be found in the supplementary material.


The direct calculation of the long term jitter requires an ensemble of $m$ pulse trains with pulse positions $\{t_n\}_m$. The timing deviations relative to an ideal pulse train are then calculated according to
\begin{align}
 \left\{ \Delta t_n = t_n - n T_{\mathrm{C}} \right\}_m,
\end{align}
where $T_{\mathrm{C}}$ is the clock time, which is defined by the sample mean $\langle T_n = t_{n+1} - t_{n} \rangle_{m,n}$ of the pulse periods. 
Characteristic for random walk, the ensemble variance $\mathrm{Var}_m \left( \Delta t_n \right)$ then grows linearly with the pulse number $n$ for sufficiently large $n$, i.e. as soon as all correlations of the dynamical system have decayed. This motivates the definition of a pulse separation dependent timing jitter \cite{LEE02c}
\begin{align}
    \sigma_{\Delta t}(n) = \sqrt \frac{\mathrm{Var}_m(\Delta t_n)}{n}.
\end{align}
The long term timing jitter is then formally obtained as the limit
\begin{align}
    \sigma_{\rm{lt}} = \lim_{n\rightarrow \infty} \sigma_{\Delta t}(n).
\end{align}
Note that the standard deviation of the pulse periods, which is known as the pulse-to-pulse timing jitter (sometimes also called cycle or period jitter\cite{LEE02c}) is given by $\sigma_{\mathrm{pp}} = \sigma_{\Delta t}(1)$.


The timing deviations $\{\Delta t_n\}$ can be described by a cumulative stochastic process 
\begin{align}
 \Delta t_n = \sum_{k=1}^n \delta T_k,
\end{align}
where the increments $\delta T_k$ are the pulse period fluctuations defined by
\begin{align}
 \left\{ \delta T_k = T_{k} - T_C = t_{k+1} - t_k - T_C \right\}.
\end{align}
While the timing deviations $\{\Delta t_n\}$ form a weak sense stationary process, the fluctuations $\delta T_k$ are bounded by the laser dynamics and thus form a stationary process, that has zero mean by definition. This also implies a zero mean for $\{\Delta t_n\}$. Hence the variance of the timing deviation reads
\begin{align}
 \rm{Var}(\Delta t_n) = \left\langle \Delta t_n^2  \right\rangle = \left\langle \sum_{k,l=1}^n \delta T_k \delta T_l \right\rangle,
\end{align}
where $ \delta T_k \delta T_l $ can be identified as the correlation function of the pulse period fluctuations. 
Since $\delta T_k$ is described by a stationary process with constant variance $\langle \delta T^2 \rangle = \langle \delta T_k^2 \rangle$, the correlation function only depends on the distance $|k-l|$ between two pulses and the variance can be written as
\begin{align}
 \rm{Var}(\Delta t_n) = \sum_{k,l=1}^n \left\langle \delta T_k \delta T_l \right\rangle = \langle \delta T^2 \rangle \sum_{k,l=1}^n \Psi_{\delta T} \left( | k - l | \right),
\end{align}
where $\Psi_{\delta T}$ is the normalized auto-correlation function of the period fluctuations. 
Rewriting the sum in terms of the distance between pulses $d = |k-l|$, the timing jitter $\sigma_{\Delta t}$ can be expressed as
\begin{align}
 \sigma_{\Delta t}^2(n) = \langle \delta T^2 \rangle \left[ 1 + 2 \sum_{d=1}^{n-1} \frac{n-d}{n} \Psi_{\delta T}(d) \right]. \label{eq:jitterfromCorr}
\end{align}
Note that delta correlated timing fluctuations, i.e. $\Psi_{\delta T}(m) = \delta_{m,0}$, lead to a pulse separation independent timing jitter. This case corresponds to a true random walk with $\langle \delta T^2 \rangle = \sigma_{\rm{pp}}^2 = \sigma_{\rm{lt}}^2$. While the authors of \cite{ELI97} use this framework to analytically analyze the consequences of a simple exponentially decaying auto-correlation function, we intend to utilize the auto-correlation function computed from sets of simulated $\{\delta T_k \}$.

However, computing $\sigma_{\Delta t}(n)$ from \eqref{eq:jitterfromCorr} up to a pulse separation of $n$ requires a good estimate of the auto-correlation function up to $n$, which can be as costly as directly computing the jitter. Nonetheless, a significant improvement of the computational cost can be achieved by fitting the auto-correlation function with a suitable model function using only small $n$.

The basic dynamic response of semiconductor lasers are relaxation oscillations between the field intensity and the carrier populations of the active media. We therefore chose a model function that is a linear combination of $j$ exponentially damped harmonic oscillations
\begin{align}
 \Psi_{\delta t}^j(n) = \sum_{k = 1}^j a_k \exp(- \Gamma_k n) \cos(\omega_k n + \phi_k) \label{eq:fitmodel}
\end{align}
with the free parameters $a_k$, $\Gamma_k$, $\omega_k$ and $\phi_k$. 

\begin{figure}[htbp]
\centering
\includegraphics[width=\linewidth]{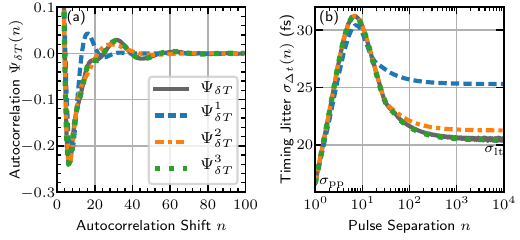}
\caption{(a) Auto-correlation function of the pulse-period fluctuations calculated from $\{\delta T_k\}$ (solid grey line) and fitted model functions $\Psi_{\delta T}^j$ for $j =1,2,3$ (dashed, dash-dotted and dotted lines). (b) Timing jitter as a function of the pulse separation calculated from $\{\Delta t_n\}$ and the fitted model functions.}
\label{fig:AC_LT}
\end{figure}

We illustrate the method in \fref{fig:AC_LT}, where (a) shows the computed auto-correlation function $\Psi_{\delta T}$ (solid grey line) and fitted model functions for $j = 1,2,3$ (blue dashed, orange dash-dotted and green dotted lines). Only the first $100\,n$ were used for the fits. We observe a fast drop of the auto-correlation function with a dip below zero, which is reproduced well by all model functions. The following relaxation to zero is characterized by further modulations, which are only captured well by the model function with $j=3$. The resulting pulse separation dependent timing jitter is shown in \fref{fig:AC_LT}\,(b). Starting with a pulse-to-pulse jitter of $\sigma_{\rm{pp}} = 16.7$\,fs, the directly computed timing jitter (grey solid line) first rises to $\approx 32$\,fs and then relaxes to its long-term value $\sigma_{\rm{lt}} = 20.5$\,fs. While the qualitative behavior is captured by all model functions, the maximum is only correctly described by $j=2,3$ and the long term value only by $j=3$. From this we conclude that our model function \eqref{eq:fitmodel} works well if $j$ is chosen properly.


\begin{figure}[htb]
\centering
\includegraphics[width=\linewidth]{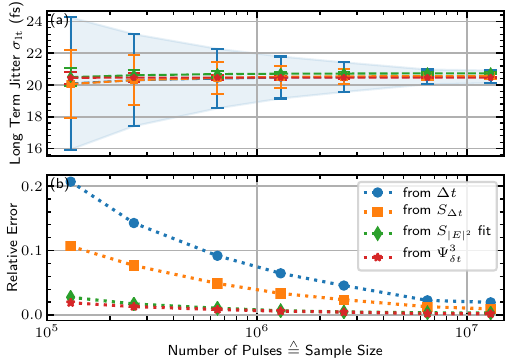}
\caption{(a) Long term timing jitter and respective standard errors as a function of the underlying number of pulses. The curves are computed from the time domain definition (blue), the von der Linde method applied to $S_{\Delta t}$ (orange), the Kéfélian method (green) and the auto-correlation method (red). (b) Relative standard errors.}
\label{fig:JitCon}
\end{figure}

As the long term timing is a statistic measure of the stochastic properties of the pulse positions, its standard error depends on the considered sample size, i.e. the number of pulses in the computed time series realizations. This is equally true for the spectral and auto-correlation based methods, where increased sample sizes generate better estimates of the power spectral density and auto-correlation function. We therefore analyze the sample size depended standard error of four methods: The direct time domain method based on $\{\Delta t_n\}$ with $n=10000$, the von der Linde method performed on the spectrum $S_{\Delta t}$, the Kéfélian method performed on the 20ths harmonic of $S_{|E|^2}$ and finally the previously introduced auto-correlation based method. Details on the sample set organization and calculation methods can be found in the supplementary.


The results are presented in \fref{fig:JitCon}, where (a) shows the computed long term jitter as a function of the sample size, with error bars indicating the standard error. Since the direct time domain method (blue) makes no assumptions about the underlying system, we expect no systematic deviations and use the resulting error range (highlighted in light blue) as a benchmark for the validity of the other methods. We find that all other methods lie within that range and thus exhibit no systematic error.

The respective relative standard errors of all four methods are additionally plotted in \fref{fig:JitCon}\,(b). We observe a $1/\sqrt{\rm{sample size}}$ dependency, i.e. the law of large numbers, for all curves, but with vastly different absolute values. While the Kéfélian and auto-correlation method produce very similar relative standard errors, the von der Linde method is worse by a factor of $\approx 5$ and the direct time domain method by a factor of $\approx 10$. In terms of computational cost (sample set size), this translates into a factor of $\approx 25$ for the von der Linde method and a factor of $\approx 100$ for the direct time domain method. 

We also report that the von der Linde method (not shown) applied to the 20ths harmonic of $S_{|E|^2}$ produces errors as low as the auto-correlation and Kéfélian method. However, the estimated value of $\sigma_{\rm{lt}}$ sensitively depends on the choice of the lower integration limit $\nu_{\rm{low}}$, which makes this method prone to systematic errors.


The difference between the methods can be explained by their underlying assumptions about the stochastic properties and their usage of the calculated time series: The direct method assumes a random walk on large time scales and only uses the last pulse position of each realization. The von der Linde method assumes a $1/\nu^2$ decay of the phase noise spectrum, which corresponds to a random walk, but integrates over a large frequency band and thus utilizes a larger number of data points. The Kéfélian methods assumes a specific lineshape of the harmonics, namely a Lorentzian, and evaluates the linewidth parameter from a large number of spectral data points, which are calculated not only from the pulse positions but from the full time-traces. Finally, the auto-correlation based method assumes an exponentially decaying oscillating pulse period auto-correlation function, where a model function has to be only fitted to small $n$, for which the estimated $\Psi_{\delta T}(n)$ is of much higher fidelity.

Based on those results, we highly recommend using either the Kéfélian or the auto-correlation method. However both methods have their specific advantages but also caveats that can lead to systematic errors in the jitter estimate: 
The Kéfélian method requires a sufficient spectral resolution, i.e. simulated time, of $S_{|E|^2}$ to successfully extract the linewidth parameter $\Delta \nu$. This problem can be somewhat mitigated by analyzing higher harmonics, but nonetheless prohibits an efficient analysis of lasers with low timing jitter. 
The auto-correlation method requires a good model function to reproduce all features of the auto-correlation function, whose existence can not by guaranteed. However, it is not limited by low jitter values and can furthermore also be applied to models that only generate pulse positions and no complete time traces \cite{DRZ13a, SCH18f}.


\begin{figure}[htbp]
\centering
\includegraphics[width=\linewidth]{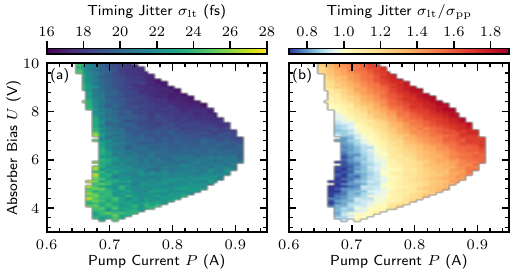}
\caption{(a) Long term timing jitter as a function of the pump current $P$ and the absorber bias $U$ computed with the auto-correlation method. (b) Long term timing jitter normalized to the pulse-to-pulse timing jitter visualizing finite time correlations. $2.5\times10^5$ Pulses have been used for each data point.}
\label{fig:LT2D}
\end{figure}

Finally, we demonstrate the strength of the auto-correlation based method by computing the long term timing jitter for the complete fundamental mode-locking regime depicted in \fref{fig:Laser}\,(a). The results are shown in \fref{fig:LT2D}\,(a). Each data point has been computed from $2.5\times10^5$ pulses, which by extrapolating the results shown in \fref{fig:JitCon} provides a relative error below $2\%$. We observe the strongest jitter with $\sigma_{\rm{lt}} \approx 28$\,fs at small pump currents and reverse biases and the smallest jitter with $\sigma_{\rm{lt}} \approx 16$\,fs at large pump currents and reverse biases.
With respect to the median, the jitter values vary by $\approx 30\%$, which in this case is only a statistically significant result due to the small relative standard error of our auto-correlation method.


The dynamics leading to the long term jitter are further highlighted in \fref{fig:LT2D}\,(b) by plotting the ratio $\sigma_{\rm{lt}}/\sigma_{\rm{pp}}$. This quantity visualizes the contribution of finite time correlation effects as they appear in $\sigma_{\rm{lt}}$ but not in $\sigma_{\rm{pp}}$.
We find that a stronger driving of the laser (both $P$ and $U$) leads to more pronounced correlation contributions (yellow/red colors), but also to a decreased pulse-to-pulse jitter $\sigma_{\rm{pp}}$ (not shown), which outweighs the correlation contributions. At $U \approx 6$\,V both effects cancel each other out leading to a long term jitter that remains constant along the pump current. At small $P$ and $U$ we also find ratios $\sigma_{\rm{lt}}/\sigma_{\rm{pp}}$ below $1.0$ (blue colors), i.e. a long term timing jitter that is smaller than the pulse-to-pulse jitter due to correlation effects.


In conclusion, we have presented a method to compute the long term timing jitter from simulated time series, which is based on a model function of the pulse-period fluctuation auto-correlation function. We benchmarked our method against already established methods by evaluating the relative standard error with respect to the computational cost and report an excellent performance. 
We attribute differences among the methods to their underlying assumptions about the system, i.e. about spectral or auto-correlation properties.
Based on our findings, we discourage the use of the direct time domain definition and give recommendations, on when to use either the very well performing Kéfélian method or our efficient auto-correlation based method.

\section*{Funding}
DFG within the Frameworks SFB 787 and SFB 910

\section*{Disclosures}
The authors declare no conflicts of interest.

\section*{Supplementary Material}
See supplementary material for data processing and calculation details regarding \fref{fig:JitCon}.

\section*{Data Availability}
The data that support the findings of this study are available from the corresponding author upon reasonable request.

\section{Supplementary Material}

\subsection{Power spectral density based timing jitter estimation methods}
Having only access to the power spectral density $S_{|E|^2}$ of the laser output, as it is the case in experiments, two established spectral methods allow for the calculation of the long term timing jitter $\sigma_{\rm{lt}}$.

The von der Linde method\cite{LIN86, KOL86} measures the integrated root mean square (rms) timing fluctuations within the offset-frequency band $[\nu_{\rm{low}}, \nu_{\rm{high}}]$ at the $h$th harmonic
\begin{align}
 \sigma_{\rm{rms}}(\nu_{\rm{low}}, \nu_{\rm{high}}, h) = \frac{T_{\rm{C}}}{2 \pi h} \sqrt{ \int_{\nu_{\rm{low}}}^{\nu_{\rm{high}}} 2 S_{\phi}(\nu_{\rm{off}}, h) \, \rm{d}\nu_{\rm{off}} }.
\end{align}
$S_{\Phi}(\nu, h)$ is the single-sided timing phase-noise power spectral density, which is defined by
\begin{align}
    S_{\Phi}(\nu_{\rm{off}}, h) = \frac{2 S_{|E|^2}(h\nu_0 + \nu_{\rm{off}})}{\int_{h\nu_0 \pm \nu_0/2} S_{|E|^2}(\nu) \, \rm{d} \nu},
\end{align}
where $\nu_0$ denotes the fundamental repetition frequency and $\nu_{\rm{off}}$ the offset frequency from the respective $h$th harmonic. Note that the factor two in the numerator is commonly not found in the literature, but is crucial for the correct estimation of the long-term jitter, since it reflects the noise contributions in the negative frequency components of $S_{|E|^2}$.

Assuming a random walk, i.e. uncorrelated timing fluctuations, the integrated rms jitter can be converted into the long term jitter according to \cite{OTT14b}
\begin{align}
    \sigma_{\rm{lt}} = \sigma_{\rm{rms}}(\nu_{\rm{low}}, \nu_{\rm{high}}) \pi \left[ \frac{1}{T_{\rm{C}}} \left( \frac{1}{\nu_{\rm{low}}} - \frac{1}{\nu_{\rm{high}}} \right) \right]^{-0.5}.
\end{align}
In practice, the assumption of uncorrelated timing fluctuations requires the integration domain to be beyond the corner frequency (induced by the lineshape), i.e. to only cover the part of the spectrum, which falls of as $1/\nu^2$. Note that assuming no correlations does not lead to the pulse-to-pulse, but to the long-term timing jitter.

Alternatively, the timing phase noise spectrum can also be computed from the sets of timing deviations $\{\Delta t_n\}$ \cite{PAS04} via
\begin{align}
    S_{\Phi}(\nu) = \left( \frac{2 \pi}{T_{\rm{C}}} \right)^2 S_{\Delta t}(\nu),
\end{align}
where $S_{\Delta t}$ is the power spectral density of $\{\Delta t_n\}$. This way, the von der Linde method can be applied if only the pulse positions are known. Moreover, the spectrum $ S_{\Delta t}$, does not obey a Lorentzian lineshape, but shows the $1/\nu^2$ divergence at small frequencies. Hence, the integration range is not restricted by the corner frequency.

The second method, introduced by Kéfélian et al.\cite{KEF08}, was developed specifically for passively mode-locked lasers and estimates the long term jitter from the repetition rate line-width $\Delta \nu_h$ of the $h$th harmonic. The jitter induced repetition rate broadening follows a Lorentzian lineshape \cite{HAU93a,ELI97}, which grows quadratically with the harmonic number $h$. Hence, the linewidth uniquely describes the timing phase noise sidebands \cite{YAM83a}, which leads to the following equation for the long term jitter via the stochastic Parseval theorem \cite{KEF08}:
\begin{align}
    \sigma_{\rm{lt}}(h) = T_{\rm{C}} \sqrt{ \frac{\Delta \nu_h T_{\rm{C}}}{2 \pi h^2} }.
\end{align}
This quantity is sometimes referred to as the pulse-to-pulse jitter \cite{KEF08,DRZ13a}. However, in this manuscript the pulse-to-pulse jitter is used to refer to the standard deviation of the pulse period fluctuations. Those standard deviations on the other hand are sometimes also referred to as the cycle jitter or period jitter \cite{LEE02c}. 

\subsection{Standard error calculation and sample set organization}

The full sample set contains 1000 time series of 10\,$\mu$s simulated time each, which corresponds to $\approx 133000$ pulses. The standard error analysis is then performed using a subsampling method, with subset sizes between one and 100. The sample size used for one timing jitter estimated is measured by the number of pulses in the respective subset. The standard error is determined from the timing jitter estimates originating from the different subsets.

The power spectra $S_{|E|^2}$ are directly calculated from the full time series using a FFT algorithm. For subsets containing multiple time series, Bartlett's method is used to average the individual power spectra.
The time series are further processed by extracting the pulse positions $t_n$, which are then subdivided into sets of 10000 for the further calculation of the timing deviations $\{\Delta t_n\}$.
The power spectra $S_{\Delta t}$ are calculated from the sets $\{\Delta t_n\}$ with a FFT algorithm and if applicable Bartlett's method for averaging. Finally, the auto-correlation functions $\Psi_{\delta T}$ are computed from the pulse period fluctuations $\{\delta T_k = t_{k+1} - t_k - T_{\rm{c}} \}$ via the Wiener–Khinchin theorem also using Bartlett's method if applicable.

In all cases, an increased sample set size improves the fidelity of the estimated quantity, i.e. the timing deviation variance $\rm{Var}(\Delta t_n)$, the power spectral densities $S_{|E|^2}$ and $S_{\Delta t}$ or the auto-correlation function $\Psi_{\delta T}$.


%

\end{document}